\documentclass[]{aastex631}

\usepackage{amsmath}
\usepackage{xcolor}
\usepackage{booktabs} 

\begin{document}

\title{Time-domain anomalies in solar and stellar flares}

\correspondingauthor{Sergey A. Belov}
\email{Sergey.Belov@warwick.ac.uk}

\author[0000-0002-3505-9542]{Sergey A. Belov}
\affiliation{Centre for Fusion, Space and Astrophysics, Department of Physics, University of Warwick, Coventry CV4 7AL, UK}

\author[0000-0002-0687-6172]{Dmitrii Y. Kolotkov}
\affiliation{Centre for Fusion, Space and Astrophysics, Department of Physics, University of Warwick, Coventry CV4 7AL, UK}
\affiliation{Engineering Research Institute \lq\lq Ventspils International Radio Astronomy Centre (VIRAC) \rq\rq, Ventspils University of Applied Sciences, Ventspils, LV-3601, Latvia}

\author[0009-0006-0995-1656]{Akshay V. Mehta}
\affiliation{Centre for Fusion, Space and Astrophysics, Department of Physics, University of Warwick, Coventry CV4 7AL, UK}

\author[0000-0002-6835-2390]{Laura A. Hayes}
\affiliation{Dublin Institute for Advanced Studies, DIAS Dunsink Observatory, Dublin D15 XR2R, Ireland}

\begin{abstract}
The temporal morphology of flare light curves encodes the underlying flare physics, and deviations from the typical flare profile may indicate the presence of mechanisms not captured by a standard flare model. To search for such time-domain deviations from a \lq\lq standard\rq\rq\ flare, we develop an unsupervised Deep Support Vector Data Description (Deep SVDD) model, which learns a compact representation of normal flares, against which unseen anomalous flares are identified.
The model is trained on synthetic light curves with a \lq\lq normal\rq\rq\ flare morphology, generated from existing analytical flare trend models with noise. Using the distribution of normal flare data, we introduce a probabilistic Flare Anomaly Index (FLAI) which allows for separating flare light curves into three distinct classes: normal data (ND), weak anomalies (WA), and strong anomalies (SA). Application of FLAI to the Kepler flare catalogue {(white light)} reveals that 36\% and 30\% of events belong to the WA and SA classes, respectively. For M- and X-class solar flares from the STIX flare list, 25\% and 32\% of events in the {15--25 keV} channel are classified as WA and SA, respectively, versus 15\% for both WA and SA classes in the {4--10 keV} channel. Thus, anomalous flares appear more frequently in the STIX high-energy channel.
These results show that both solar and stellar flares often deviate from the normal flare population used for model training, suggesting departures from the standard flare scenario, such as modified energy release and dissipation, or the development of wave and oscillatory processes in flare sites.
\end{abstract}


\section{Introduction} \label{sec:intro}
Solar and stellar flares are bursts of electromagnetic energy emission, which are observed from radio to X-ray and gamma-ray bands \citep{Kowalski2024}. Compared to stellar flares, solar flares allow for spatially resolved observations \citep[see e.g. reviews by][]{Fletcher2011, Benz2016}, which provide insight into the physical processes behind them \citep{Shibata2011}. The possibility to observe solar flares with spatially resolved, multi-wavelength data across multiple spatial and temporal scales led to the development of the standard solar flare model \citep{carmichael196454, sturrock1966model, hirayama1974theoretical,kopp1976magnetic}. It is commonly believed that the physical processes involved in this model are also responsible for stellar flares, and that the energy release in both solar and stellar flares is driven by magnetic reconnection \citep{Shibata2011}. Several observational findings support this interpretation. For example, the empirically established Neupert effect, which links thermal and non-thermal flare emissions \citep{Neupert1968}, has also been observed in stellar flares \citep{Hawley1995, Gudel1996, 2010ARA&A..48..241B, 2021ApJ...912...81K}. {However, some studies demonstrated that a significant fraction of both solar \citep[20\% in][]{Dennis1993} and stellar \citep[65\% in][]{Tristan2023} flares does not follow the Neupert effect. }

Solar and stellar flare light curves usually exhibit a \lq\lq standard\rq\rq\ profile consisting of a fast rise phase followed by a slow decay, most pronounced in thermal flare emission. While this description is often applied to non-thermal bands, their light curves generally show distinct morphological features (sharper rises, more frequent multiple peaks, and shorter durations) due to their different driving mechanisms. The temporal morphology of flare light curves encodes the plasma dynamics occurring in flares: the rise phase reflects magnetic energy release via reconnection \citep{Shibata2011} and the consequent plasma heating, while the decay phase represents plasma cooling due to radiation and thermal conduction \citep[see e.g.][]{1992A&A...253..269J, 2007A&A...471..271R, 2025ApJ...987L...9B}. \cite{Gryciuk2017} derived an elementary flare profile as a convolution of Gaussian heating and exponential cooling, and applied it to solar soft X-ray emission data observed by the Solar Photometer in X-rays (SphinX) on the CORONAS–PHOTON mission. \cite{Davenport2014} used white-light data with a 1-minute cadence from Kepler to collect over 6100 flare events occurring on a single M dwarf during 11 months. Among these, 885 flares were identified as having a \lq\lq standard\rq\rq\ profile and were used to generate a median flare template. The decay phase of this template can be decomposed into two consecutive exponential cooling phases. These phases are commonly attributed to dominant cooling by radiation and thermal conduction, respectively \citep{2005psci.book.....A}.

A similar approach was applied to Sun-as-a-star observations by \cite{Kashapova2021} using the data from the Atmospheric Imaging Assembly (AIA) instrument onboard the Solar Dynamics Observatory (SDO). In this study,  single-peak flare light curves with smooth decay phases were selected. As a result, the dataset contained 102 flare events seen in the 1600\,\AA\ and 304\,\AA\ channels, and 54 events from the 1700\,\AA\ channel. The decay phase of the selected flares was found to be best described by a two-exponential decay model, similar to that used for the M-dwarf star. However, for the 1700\,\AA\ channel, the decay was slower than that observed for the M dwarf, which led the authors to conclude that the observed M-dwarf emission originated from a denser atmospheric layer than the solar flare emission detected in the 1700\,\AA\ channel. \cite{Motyk2022} applied analytical expressions describing temperature decay due to either thermal conduction or radiative losses from \citep{Cargill1995} to both the solar \citep{Kashapova2021} and stellar \citep{Davenport2014} datasets in order to extract information about the dominant cooling processes. The study revealed that radiation dominates the cooling phase for almost all spectral bands, except for the 304\,\AA\ channel, where radiative cooling begins to dominate earlier than predicted by the standard model. A comparison between the solar and M-dwarf datasets suggested similarities in the cooling processes operating in solar and stellar flares and their dependence on atmospheric structure.
In contrast, \citet{2021ApJ...912...81K} found signatures of a faster (in comparison with the solar case) cooling of the soft X-ray-emitting plasma in stellar flares, by analysing flare time profiles in Kepler and XMM-Newton data.
In microwaves, \cite{Motyk2025} analysed 116 flare times profiles observed with the Siberian Radioheliograph in the frequency range of 3--24\,GHz. The broadband nature of the data enabled the construction of median microwave flare templates for emission from optically thick and optically thin sources. Both templates are found to be almost identical, possibly indicating a dominant contribution from accelerated electron precipitation. {In addition to analysing flare decay profiles and template morphologies, flare light curves can also be characterised using integral parameters such as the impulsiveness index \citep{Kowalski2013, Tamburri2024}, defined as the ratio between the flare intensity amplitude and its half-width time. In particular, a high impulsiveness index was shown to correlate with a high maximum reconnection rate \citep{Tamburri2024}.}

Observed flare light curves often deviate from the mean shape predicted by the standard flare model. In particular, modelling the observed duration of the impulsive flare phase ($10^1$--$10^3$\,s), during which energy is rapidly released through magnetic reconnection, often requires anomalously high values of electric resistivity, manifesting the long-standing problem of fast magnetic reconnection \citep[see e.g.][for comprehensive reviews]{2009ARA&A..47..291Z, Shibata2011, 2022LRSP...19....1P}. Another feature common to both solar and stellar flare observations is the pronounced quasi-periodic modulation of light curves across multiple electromagnetic bands, commonly referred to as quasi-periodic pulsations \citep[QPP, see e.g.,][for reviews]{2009SSRv..149..119N, 2010PPCF...52l4009N, 2016SoPh..291.3143V, 2020STP.....6a...3K, 2021SSRv..217...66Z}. While not described by the standard flare model, QPP can provide valuable seismological information about plasma processes occurring in flares. At least fifteen mechanisms have been proposed to explain QPP formation \citep{2018SSRv..214...45M}, varying from self-induced repetitive reconnection  to modulation of the magnetic reconnection rate or the parameters of the emitting plasma by magnetohydrodynamic (MHD) waves. These mechanisms are expected to produce QPP with distinct characteristic time signatures and parameters, which can serve as observational imprints of the underlying physical processes operating in flare sites.

Thus, both small- and large-scale variability in flare light curves carry important diagnostic information about the underlying plasma processes and can therefore be used to constrain and improve existing flare models.
Detecting such deviations from the expected standard scenario in flare time-domain observations often requires the use of sophisticated data analysis techniques \citep[see e.g.][]{Broomhall2019, 2022SSRv..218....9A}, which significantly limits the scale and efficiency of statistically robust studies.
This motivates the development of automated techniques capable of efficiently mining large flare datasets. Among the available approaches, machine-learning-based methods show particular promise. In particular, neural networks can be implemented to extract flare light curves from stellar time-series data \citep{Vida2021, Jia2024}. The features extracted by convolutional neural networks can be used to study flare time-domain morphological properties of flares and connect these properties with observed coronal mass ejections \citep{Tan2025}. Moreover, ML techniques are capable of finding small-scale features on top of flare light curves, in particular, a fully convolutional neural network was recently used to extract rapidly decaying harmonic QPP from solar and stellar observations \citep{Belov2024, Wang2025}.

In this work, we aim to identify solar and stellar flares deviating from a \lq\lq typical\rq\rq\ flare profile, i.e. containing statistical time-domain anomalies. To achieve this goal, we train neural-network model, which learns a compact representation of a normal-flare data model, against which unseen anomalous flares are identified. The paper is organised as follows. In Section \ref{sec:data}, we describe the concept of a \lq\lq normal\rq\rq\ flare and the dataset we used in this study to represent it. While flare light curve morphology differs across emission types, all broadly share a common impulsive rise and slower decay profile \citep{Fletcher2011, Benz2016}, which motivates the use of the same normal flare population across the solar and stellar datasets considered in this study. Emission and instrument-specific differences are discussed in Sections \ref{sec:data} and \ref{sec:real-data}.
In Section \ref{sec:model}, we introduce the model architecture to detect time-domain anomalies. In Section \ref{sec:real-data}, we introduce a Flare Anomaly Index (FLAI) as a quantitative measure to statistically characterise deviation of a given flare from the \lq\lq normal\rq\rq\ data, and apply it to stellar and solar flare data from Kepler and the Spectrometer Telescope for Imaging X-rays (STIX) on Solar Orbiter, respectively.
Section \ref{sec:meanprofile} discusses the average flare templates and data variability for normal and anomalous flares.
We draw our conclusions and describe the deployment of the model as a browser-based application with a user-friendly interface in Section \ref{sec:concl}.

\section{Dataset} \label{sec:data}
To extract anomalous flare light curves from input time-series data, it is first necessary to establish what constitutes a \lq\lq normal\rq\rq\ flare and how such events are represented in the data. In other words, anomaly detection requires a model of the normal flare population against which deviations can be identified. In this work, we reuse the open-access dataset of synthetic flare light curves presented in \citet{qpp_dataset} and previously employed by \citet{Belov2024} to train a fully convolutional neural network (FCN) for detecting QPPs in solar and stellar flares. The choice of the synthetic data allows us to control its quality and ensure that it does not contain any undesirable features. The dataset consists of two subsets: one containing flare light curves with noise of randomly varying amplitude and one containing noise-free flare light curves. Each subset comprises 90,000 synthetic flare light curves, each consisting of 300 data points, generated using the available analytical flare trend models of \citet{Davenport2014}, \citet{Gryciuk2017}, and \citet{Broomhall2019}, with 50\% of the samples additionally containing exponentially decaying harmonic QPP signals. The adopted flare trend models include both solar and stellar light-curve representations. The template proposed by \citet{Davenport2014}, which captures white-light emission from M dwarfs, has also been applied to solar flares \citep{Kashapova2021}. The second model \citep{Gryciuk2017} is derived for solar soft X-ray observations, while the third model \citet{Broomhall2019} is an empirical formulation combining two half-Gaussian functions to describe the rise and decay phases. Together, this set of models enables us to represent both solar and stellar flare light curves. For training our anomaly detection model, we use the other half of the noise-free subset that contains only flare trends and no QPPs, yielding a total of 45,000 synthetic flare light curves. Using the noise-free subset allows noise to be added separately in a controlled manner, with the noise-to-signal ratio ($NSR$) treated here as a free parameter. These light curves therefore define the \lq\lq normal\rq\rq\ flare population used to train our model. The remaining unseen subset with QPPs is used for initial validation, where the QPP signatures serve as representative examples of potential time-domain anomalies that the model is designed to detect. We split our synthetic flare data into training (80\%), validation (10\%), and test (10\%) sets.

\begin{figure}[h]
\centering
\includegraphics[width=0.8\textwidth]{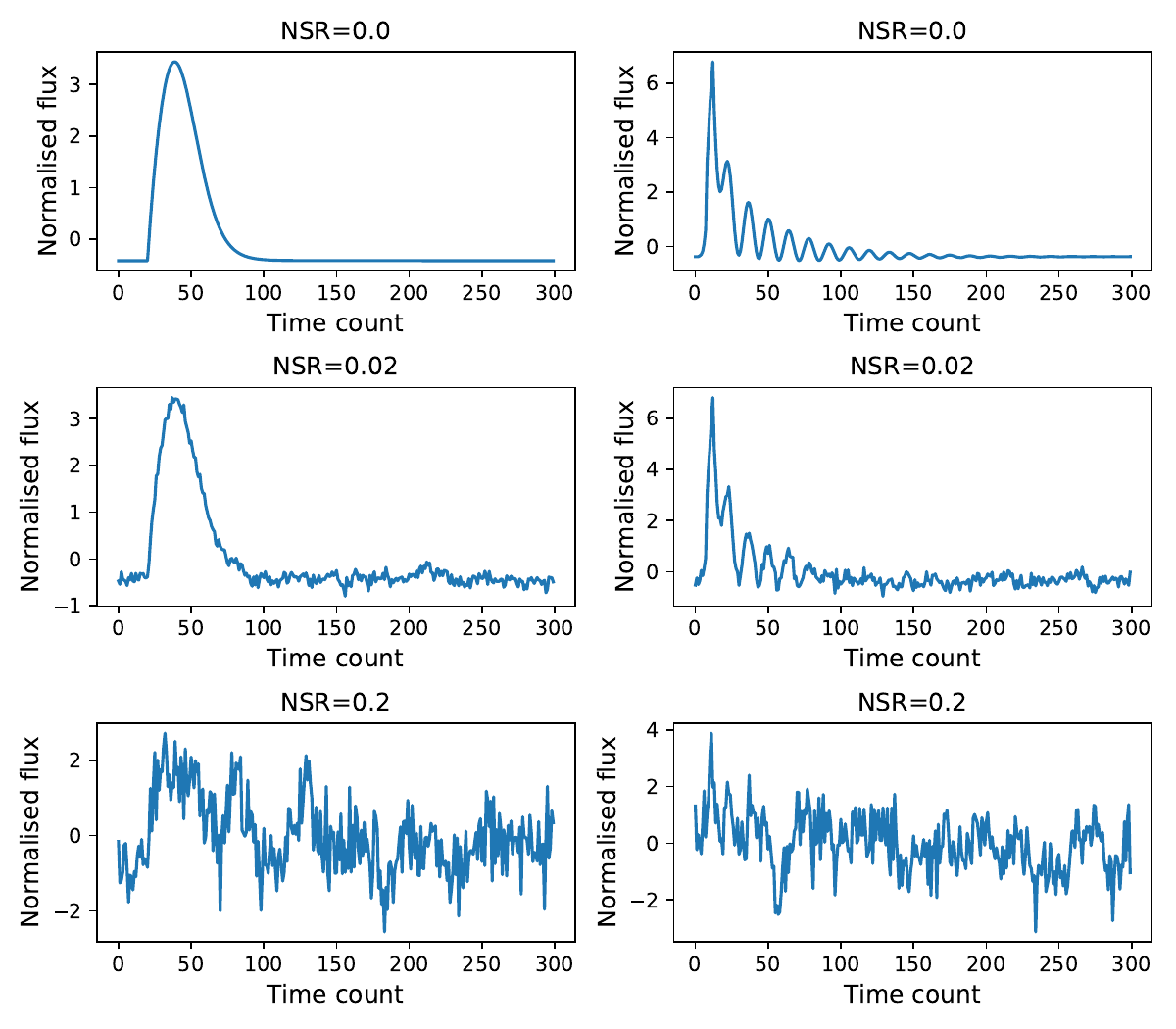}
\caption{Examples of synthetic light curves with imposed noise. The left column represents a flare trend only, while the right column shows a light curve with both flare trend and a QPP. Each light curve is standardised using its own mean and standard deviation.\label{fig:data_noise}}
\end{figure}

As solar and stellar flare light curves are typically contaminated by red and white noise \citep{Inglis2015, Kolotkov2018, Hayes2020}, their combination is added to the flare trend samples during training.  As a result, the network is exposed to slightly different data at each epoch. For each noise component, we use equal amplitudes  $A_N=NSR\times A_\mathrm{flr}$, where $A_\mathrm{flr}$ is the flare amplitude with respect to a background level, and $NSR$ is the noise-to-signal ratio. Fig.~\ref{fig:data_noise} shows an example of a \lq\lq normal\rq\rq\ flare light curve and a flare light curve with QPP as an anomaly example, with added noise with three different values of $NSR$: 0.0, 0.02, 0.2. It is evident that, at higher $NSR$ values, distinguishing a \lq\lq normal\rq\rq\ light curve containing a flare trend only becomes increasingly difficult. This implies that the range of $NSR$ values used during training must be constrained accordingly.

\section{Anomaly detection model} \label{sec:model}
In this work, we use Deep Support Vector Data Description \citep[Deep SVDD,][]{pmlr-v80-ruff18a} to separate anomalous flares from normal data. Under this approach, a feature extraction model (feature extractor) is trained to map normal input data into a hypersphere of radius $R$ in an $M$-dimensional feature space. Consequently, unseen anomalous data are expected to lie outside this hypersphere. To achieve this, the following loss-function is used:
\begin{equation}
\label{eq:Loss_f}
\mathcal{L} = R^2 +
\frac{1}{\nu N}
\sum_{i=1}^{N}
\max\left(
\sum_{j=1}^{M} (f_{ij} - c_j)^2 - R^2,\; 0
\right),
\end{equation}
where $N$ is the number of data samples in the batch/training dataset, $\nu$ is the regularisation parameter, $f_{ij}$ is the $j$-th component of the output feature vector for sample $i$, and $c_j$ is the $j$-th component of the hypersphere centre, defined as the centre of mass of the feature vectors produced by the model before training begins. The parameter $\nu$ controls the fraction of normal data allowed to lie outside the hypersphere, thereby introducing a soft decision boundary. This soft boundary prevents the hypersphere from over-expanding during training. Using this parameter, the hypersphere radius $R$ is determined as $1-\nu$ percentile of the distances between feature vectors and the hypersphere centre. In this study, we use $\nu=0.1$, which ensures that approximately 90\% of our normal data used for training lie within the hypersphere, and 10\%  may appear outside it. {The value $\nu=0.1$ was recommended by the Deep SVDD author\footnote{ https://github.com/lukasruff/Deep-SVDD/issues/7} and is used in other works \citep[see e.g.][]{Jiang2023}.} The discussed method is an example of an unsupervised learning approach, in which the model is trained without explicit labels and instead learns the underlying structure of the data, in contrast to previous studies \citep{Vida2021, Jia2024, Belov2024, Tan2025} that rely on supervised training with labeled examples.

\begin{figure}[h]
\includegraphics[width=\textwidth]{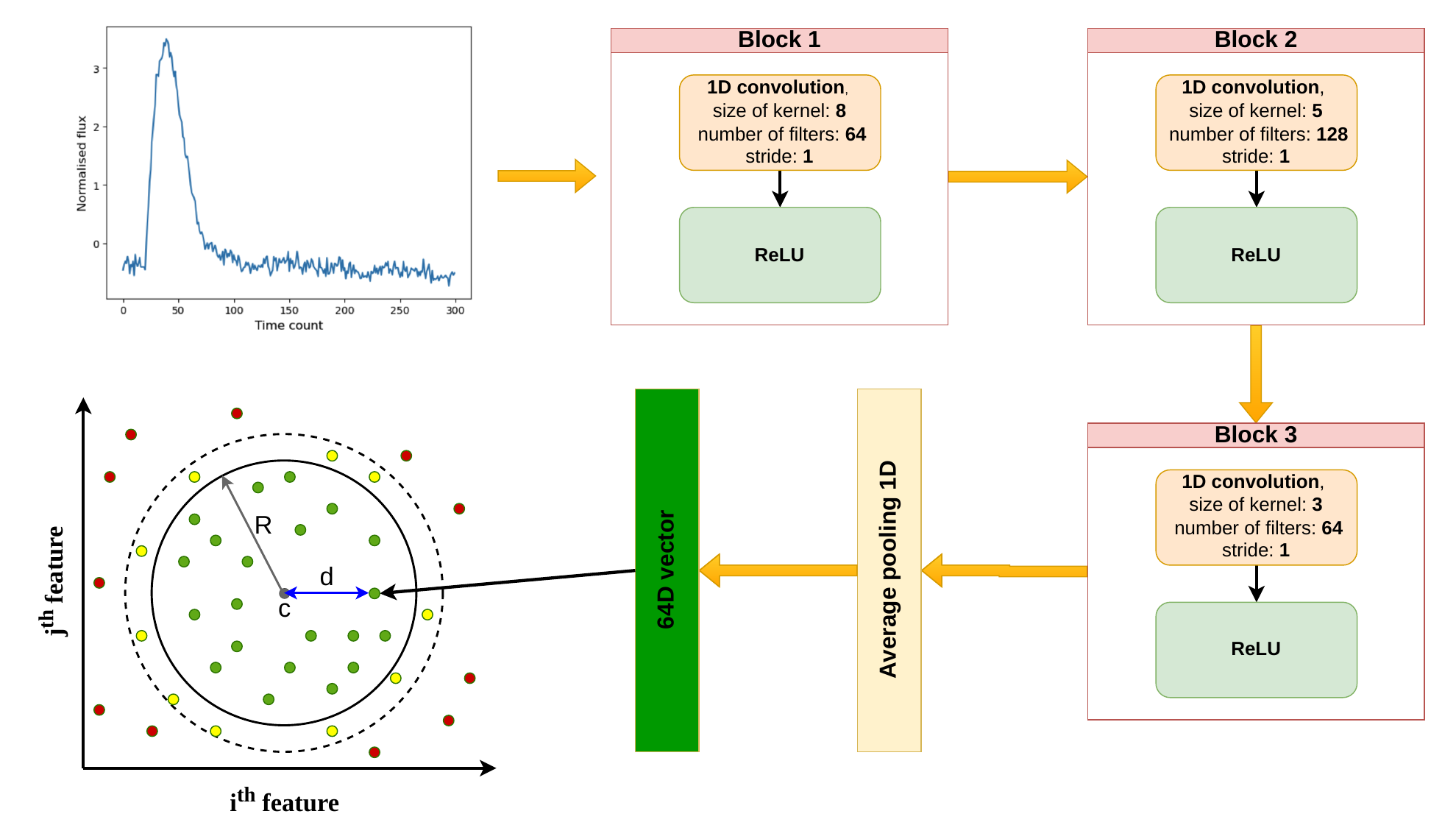}
\caption{Architecture of the Deep SVDD model for anomaly detection. Upper-left panel: an example output light curve. Lower-left panel: a schematic 2D cross-section of the hypersphere (solid circle) and the feature vectors produced by the model (coloured points). Green, yellow, and red points denote normal data within the hypersphere, $\sim$10\% of the training data allowed to lie outside the hypersphere due to the soft boundary condition, and anomalous data, respectively.\label{fig:sheme}}
\end{figure}

As a feature extractor, we use the FCN architecture proposed by \cite{Wang2017} and later adopted by \cite{Belov2024} to detect exponentially decaying harmonic QPPs in flare light curves. The model architecture is shown in Fig.~\ref{fig:sheme}. The main difference from the original architecture is that, for the anomaly detection task using Deep SVDD, batch normalisation layers within convolutional blocks are removed, as well as the fully connected layer used for mapping feature vectors to class labels. In this study, we use $M=64$, which means the model derives 64 characteristics for each flare light curve. As a result, the model, when fed with a normal flare, produces a 64D feature vector that lies within a hypersphere. This ensures that the latent representation is compact enough to avoid overfitting but still capable of capturing enough information from the input data. Therefore, every light curve can be described by the distance $d$ between a corresponding feature vector and the hypersphere centre:
\begin{equation}
\label{eq:d2}
    d^2_i = \sum_{j=1}^{M} (f_{ij} - c_j)^2.
\end{equation}
The lower-left panel of Fig.~\ref{fig:sheme} shows a schematic representation of a 2D cross-section of the hypersphere and the feature vectors produced by the model. The green points inside the solid circle of radius $R$ represent normal data mapped within the hypersphere. The space between the solid and dashed circles contains the remaining 10\% of the data, which are allowed to lie outside the hypersphere due to the soft boundary condition. This population is denoted by yellow points and may include both normal and weakly anomalous data. The red points with $d/R \gg 1$ indicate clearly anomalous samples. For the sake of simplicity, we do not add any additional input channels to the model, using only standardised light curves as input.

\begin{figure}[h]
\centering
\includegraphics[width=0.8\textwidth]{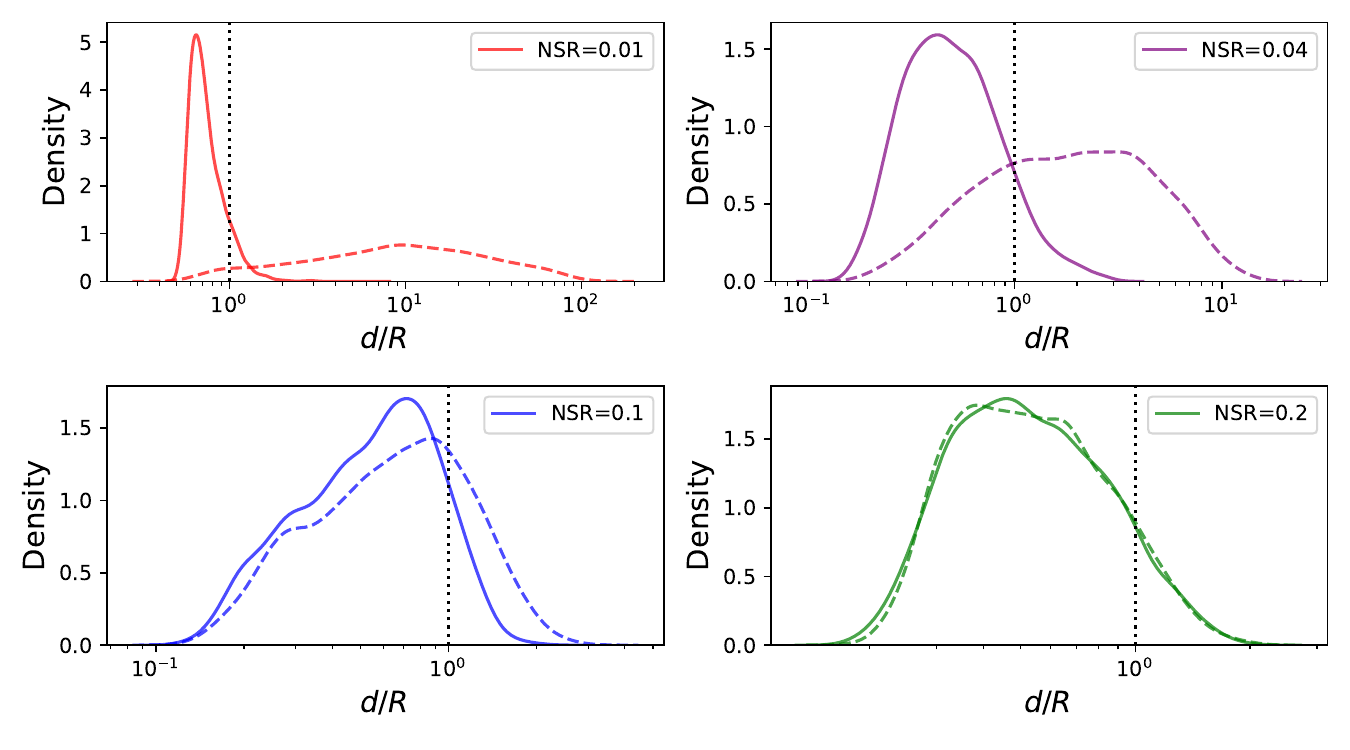}
\caption{The distribution of the distance measured in $R$ between data points in an $M$-dimensional feature space and the centre of the hypersphere for normal data (solid lines) and anomalous data containing QPPs (dashed lines). The dotted vertical line denotes the sphere boundary (spanning 90\% of normal data).}
\label{fig:train_dist}
\end{figure}

To test the model's capability to separate \lq\lq normal\rq\rq\ data from data having different properties (anomalies), we trained our anomaly detector on the normal data described in Section \ref{sec:data} for 20 epochs. We conducted several training and testing runs for different $NSR$ values, starting from 0.0 (no noise). During the training stage, the noise was generated and added dynamically; thus, the network did not encounter exactly the same light curves in each epoch. At the test stage, we used two test sets: one similar to the training set (normal) and one containing QPP signals. The noise was also added dynamically to these subsets. Figure~\ref{fig:train_dist} illustrates the distribution of the $d/R$ parameter (the normalised distance between data points and the hypersphere centre) for normal data (solid lines) and anomalous data containing QPPs (dashed lines). The top left panel of Fig.~\ref{fig:train_dist} shows that, for the case of weak noise ($NSR=0.01$), the datasets with flares only and flares with QPPs can be statistically separated. At the same time, it is clearly seen in the other panels of Fig.~\ref{fig:train_dist} that an increase in noise amplitude causes the two distributions to converge together, making them almost indistinguishable at $NSR=0.2$. This outcome is expected, as high noise levels make the separation of time-series features more challenging (see e.g. the bottom row in Fig.~\ref{fig:data_noise}), even by visual inspection.

To create the final model, we train the network for 20 epochs, adding noise dynamically with $NSR$ values uniformly sampled between 0.01 and 0.1, based on the results shown in Fig.~\ref{fig:train_dist}. The resulting model is then used to identify anomalous flare light curves in real observational data.

\section{Real-data anomalies} 
\label{sec:real-data}

\subsection{Flare Anomaly Index (FLAI)} 
\label{sec:flai}
Before we apply the developed anomaly detector to real data, it is instructive to discuss how anomalies can be quantified and measured. If a light curve results in a feature vector lying inside the hypersphere, i.e. $d < R$, it is classified as part of the normal data population. However, the condition $d > R$ does not necessarily imply that a given light curve is anomalous, because, in this study, we use a Deep SVDD model with soft boundaries, allowing a fraction $\nu$ of normal data to lie outside the hypersphere. This prevents hypersphere inflation and enables the model to learn a compact representation of the core normal data distribution. Consequently, the simple distance measure $d/R$ is not sufficient as an anomaly score, since a non-negligible fraction of normal data lies outside the hypersphere. This issue can be addressed statistically by fitting the tail of the normal data distribution ($d > R$) with a smooth function, such as a generalised Pareto distribution (GPD). Once obtained, the fit can be used to calculate the probability that a given flare profile, with distance $d$ in the $M$-dimensional feature space, belongs to the tail of the normal data distribution, $P_{\mathrm{tail}}$ (i.e. to the fraction of normal data permitted outside the hypersphere). A high value of $P_{\mathrm{tail}}$ indicates that the flare is consistent with the normal data distribution, whereas a low value suggests that the flare is likely anomalous.

To implement the approach explained above, we calculated the distances $d$ for our training data and an estimated hypersphere radius $R$ calculated as the (1-$\nu$)-percentile (90\% in this study) of the normal data distances. Then, the tail of the distribution was fitted with a GPD function using {the genpareto.fit function\footnote{https://docs.scipy.org/doc/scipy-1.17.0/reference/generated/scipy.stats.genpareto.html} from the stats module in the SciPy Python package.} It allows us to estimate the probability that a given value of $d$ belongs to the tail of the distribution, as follows:

\begin{equation}
\label{eq:pareto_sd}
P_{\mathrm{tail}}=P\left(Y>y\right)=\left(1 + \frac{\xi y}{\beta}\right)^{-1/\xi},
\end{equation}
where $y = d - R$, and $\xi$ and $\beta$ are the shape and scale parameters of the GPD. {It should be noted that the use of the GPD function is not unique in this context, and alternative decaying functions (e.g. polynomial functions), as well as direct numerical integration, may also be used to estimate $P_{\mathrm{tail}}$.}

\begin{figure}[h]
\centering
\includegraphics[width=\textwidth, trim=3cm 0cm 3cm 0cm]{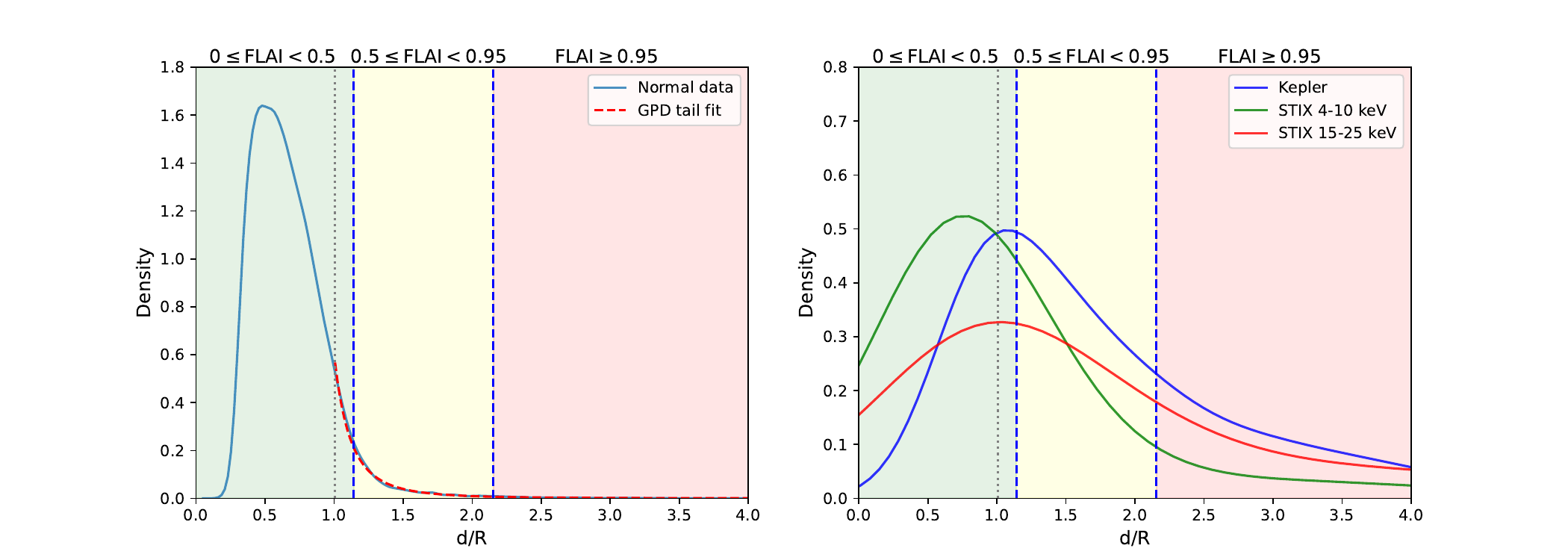}
\caption{Left panel: the data distribution with $d/R$ for normal data used for the network training (blue curve) and the GPD tail fit (dashed red curve). Right panel: the data distribution with $d/R$ for Kepler data (blue curve) and STIX low-energy (green curve) and high-energy (red curve) bands. The grey dotted line represents a threshold equal to the 90th percentile of the normal data, while the blue dashed lines indicate $P_{\mathrm{tail}} = 0.5$ and $P_{\mathrm{tail}} = 0.05$. The green, yellow, and red shaded regions represent the regions corresponding to the ND, WA, and SA classes, respectively.
}\label{fig:d_dist}
\end{figure}

The left panel of Fig.~\ref{fig:d_dist} shows the distribution of the normal data used for training in terms of $d/R$, along with its tail fitted by the GPD.
The value of $P_{\mathrm{tail}}$ can be used to construct a new Flare Anomaly Index (FLAI),
\begin{equation}
    \mathrm{FLAI}=1-P_{\mathrm{tail}}.
    \label{eq:flai}
\end{equation}
FLAI allows us to separate flare data into the following classes (see also the shaded regions in Fig.~\ref{fig:d_dist}):

\begin{itemize}
    \item Normal data (ND): $0.0\le \mathrm{FLAI}<0.5$; 
    \item Weak anomalies (WA): $0.5\le\mathrm{FLAI}<0.95$; 
    \item Strong anomalies (SA): $\mathrm{FLAI}\geq0.95$. 
\end{itemize}

Our reasoning for choosing the anomaly class boundaries is as follows. First, $\mathrm{FLAI}=0.5$ corresponds to the point in the feature space at which it is no longer possible to reliably discriminate whether a data point belongs to the tail (i.e. the flare has an equal 50\% probability of belonging to the tail of the normal population or not), which makes it a good estimate of the normal data boundary. The value $\mathrm{FLAI}=0.95$ covers almost the whole tail meaning that flares with $\mathrm{FLAI}\le0.95$ can be considered significantly anomalous.

\begin{figure}[h]
\centering
\includegraphics[width=\textwidth]{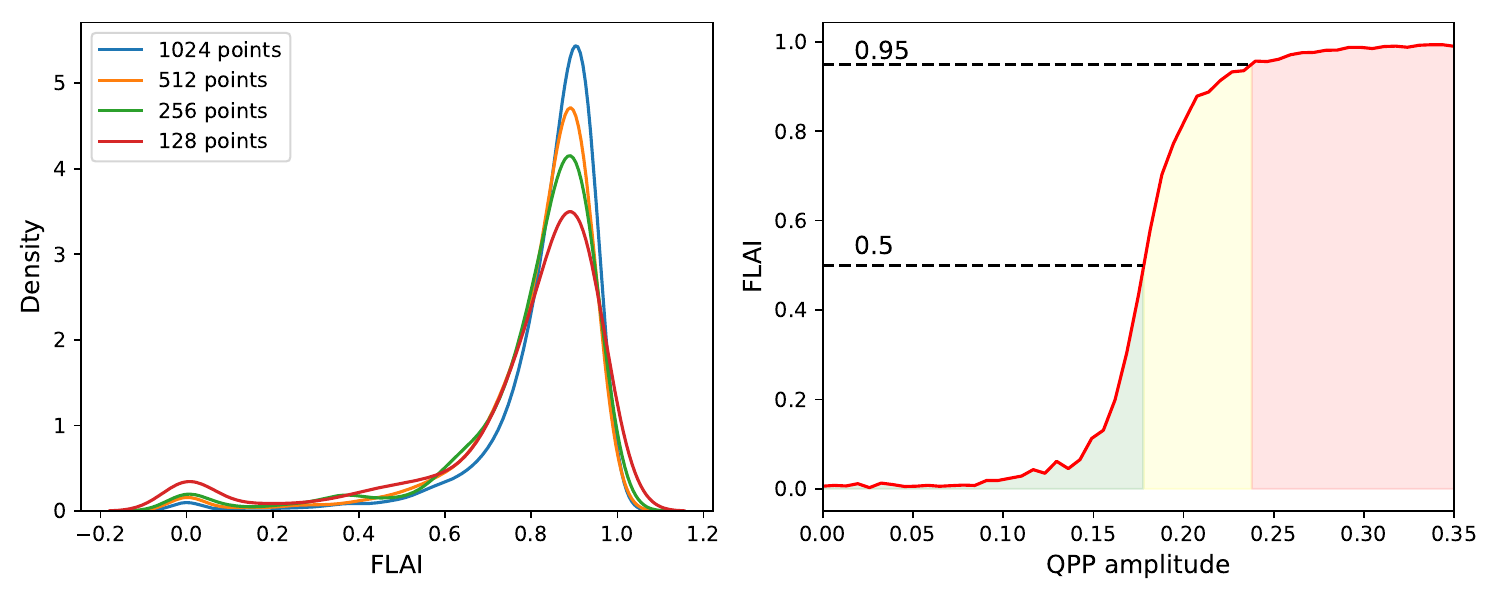}
\caption{{Left panel: distributions of the FLAI values obtained for synthetic flare time series with different temporal resolutions corresponding to 1024, 512, 256, and 128 data points. Each distribution was produced from 1000 noise realisations with $NSR=0.02$, where the original flare trend was combined with a damped QPP signal of amplitude $a_0 = 20\%$ of the flare amplitude. All time series were subsequently resampled to 300 points before being passed to the Deep SVDD model. Right panel: the dependence of the average FLAI on the QPP amplitude $a_0$ for synthetic time series containing 300 data points. For each amplitude value, 500 noise realisations were generated with randomly selected $NSR$ values between 0.01 and 0.1. The red curve shows the mean FLAI value as a function of the QPP amplitude. The green, yellow, and red shaded regions correspond to the ND, WA, and SA classes, respectively.}
}\label{fig:f_test}
\end{figure}

{Before applying FLAI to real data, it is instructive to examine its sensitivity using a series of synthetic tests. For these tests, we adopted a fixed flare trend based on the flare profile model of \citep[][see Eq.~(4) therein]{Gryciuk2017}, with parameters $B=1.0$, $C=2.5$, and $D=0.5$ providing a flare with clear rise and decay phases. The flare spans the interval between 0.25 and 1.0 in normalised time units, where the total time-series duration is 1.0. To introduce anomalous behaviour, we added an exponentially decaying harmonic QPP on top of the trend starting from its peak in a form:
\begin{equation}
    \mathrm{QPP}\left(t\right)=a_0\exp\left(-\frac{t-t_0}{\tau}\right)\cos{\left(\frac{2\pi\left(t-t_0\right)}{P}\right)},
    \nonumber
\end{equation}
where $t_0$ is the oscillation start time chosen to coincide with the flare peak; $a_0$, $P$, and $\tau$ are the QPP amplitude, period, and damping time, respectively. In this experiment, we set $P$ to one-sixth of the flare duration ($3/24$ in normalised time units) and $\tau = 3P$. In our first test, we explore how FLAI is sensitive to data resolution and resampling to a fixed data length of 300 points. For this test, we generated time series consisting of 1024 data points by combining the flare-trend realisation described above with a QPP component whose amplitude $a_0$ was set to 20\% of the flare amplitude. Then, in every run, we add noise on top of this time series with $NSR=0.02$ (see Section \ref{sec:data}), and produce coarser time series by degrading its resolution by 2, 4, and 8 times, respectively. As a result, for each noise realisation, we have 4 time series with lengths of 1024, 512, 256, and 128 points, respectively. We repeat this procedure 1000 times to produce 4000 time series in total. Finally, this data was fed into the Deep SVDD model, where it was either under- or over-sampled to 300 data points. The left panel of Fig.~\ref{fig:f_test} shows the resulting FLAI distributions for each time-series length. In this test, FLAI is found to demonstrate weak sensitivity to time-series resolution and resampling, as the obtained distributions show pronounced peaks near $0.9$ for all resolutions; however, decreasing the resolution leads to lower peak values, broader distributions, and the appearance of samples classified as normal ($\mathrm{FLAI}\approx0$).}

{In the second test, we aim to quantify the proposed boundaries for the ND, WA, and SA anomaly classes and estimate what FLAI is produced by anomalies with different amplitudes. To proceed with it, we fixed the data length to 300 points to avoid the influence of resampling on the result. We used the same flare-trend and QPP models as in the first test, while varying the oscillation amplitude $a_0$ from 0 to 35\% of the flare amplitude. For each amplitude, we generate 500 time series with different noise realisations. The corresponding $NSR$ values were randomly selected between 0.01 and 0.1 to be consistent with the training noise distribution. Next, we calculated the FLAI value for each time series and computed the average FLAI for every amplitude level. The results are shown in the right panel of Fig.~\ref{fig:f_test}, where the red line illustrates the dependence of the mean FLAI on the QPP amplitude. The green, yellow, and red shadings denote the ND, WA, and SA regions, respectively. As expected, the average FLAI increases with increasing oscillation amplitude and crosses the WA-class threshold of 0.5 at approximately $a_0 \approx 0.18$. This value is consistent with the highest noise amplitude ($NSR=0.1$, see Sec.~\ref{sec:model}) present in the training dataset, indicating that the model interprets lower-amplitude oscillations as insignificant compared to the noise levels encountered during training. To reach the SA class, the QPP amplitude must exceed the noise threshold by a factor of $\sim 2.3$.}

\subsection{Kepler flares}\label{sec:kepler}

To apply FLAI to real data, we start from stellar flare light curves observed by Kepler with a 1-min cadence. We use a dataset comprising 2,274 light curves presented in \citet{Belov2024} and available on Harvard Dataverse \citep{kepler_dataset}, which was obtained from a larger Kepler flare catalogue \citep{Balona2015} by detrending the light curves with a parabolic fit to remove the effects of stellar rotation and selecting the best-fitted samples based on the $\chi^2$-criterion. Additionally, the light curves are interpolated to ensure that each time series contains 300 data points to be consistent with the training time-series length. {It should be mentioned that the 1-minute cadence of the Kepler observations may not fully resolve rapid variability during the rise phase \citep{Howard2022}. In particular, short-timescale flare structure may be smeared out, which can affect the observed flare morphology and the anomaly properties inferred by the model.}

The right panel of Fig. \ref{fig:d_dist} shows how the stellar flare data obtained from Kepler are distributed in terms of $d/R$ (blue curve). It can be seen that the Kepler data distribution extends significantly beyond the characteristic boundaries of the normal data distribution (vertical dashed lines) derived from training, which indicates a substantial fraction of Kepler flare data deviating from the normal flare population. This suggests that the representation of a normal flare model used to construct our training dataset cannot fully explain the time variability of flare properties observed in Kepler data. There are multiple possible sources of such deviations. First, the Kepler flare samples can structurally differ from our \lq\lq normal\rq\rq\ flare model, e.g. by containing series of flares or QPP, which may indicate additional underlying physical processes. Also, the light curves may contain artefacts resulting from data pre-processing (e.g. imperfect stellar rotation trend removal). Differences in noise characteristics in our \lq\lq normal\rq\rq\ flare model and in Kepler data, including higher noise amplitude in observations, can also contribute to these deviations.

\begin{figure}[h]
\centering
\includegraphics[width=0.8\textwidth]{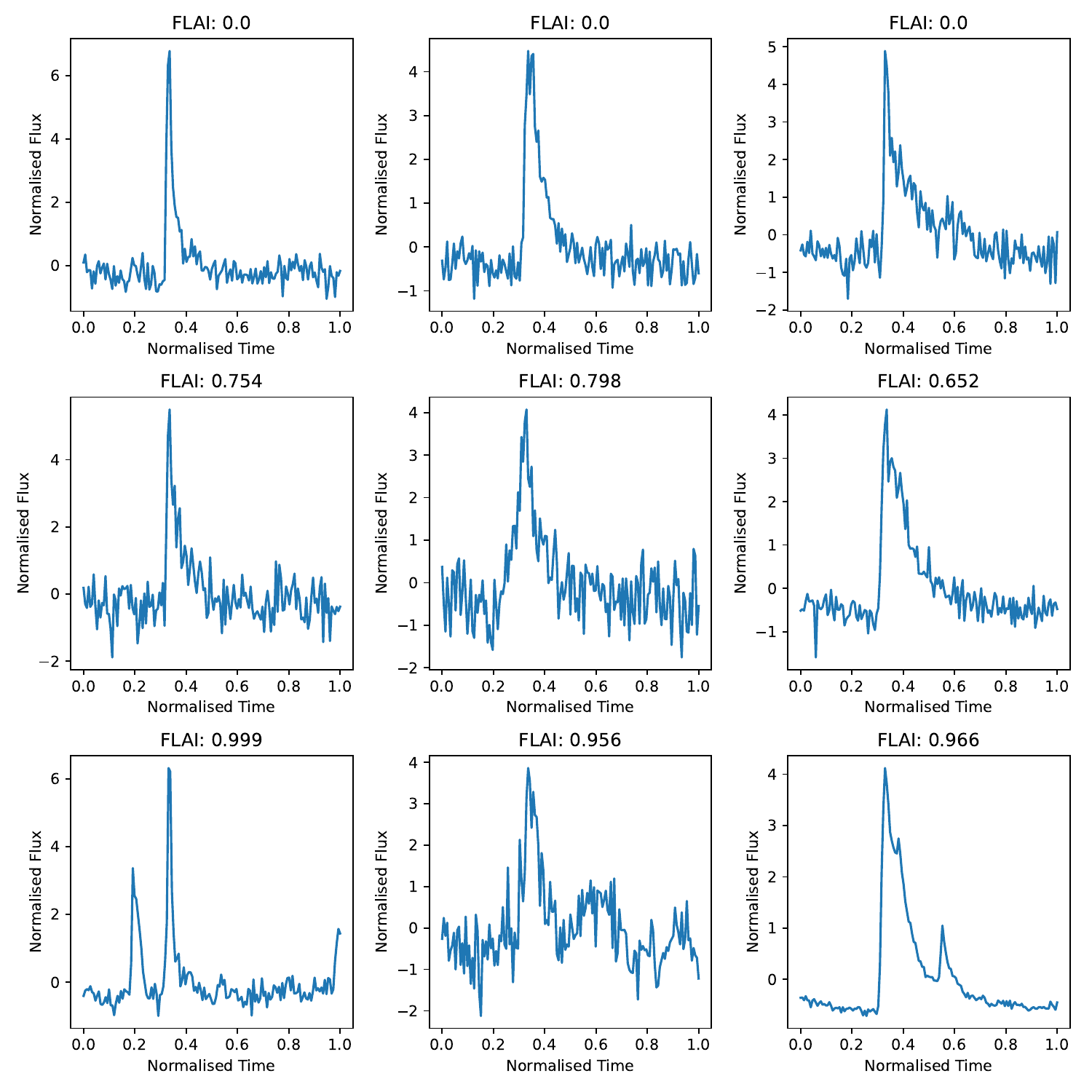}
\caption{Examples of Kepler light curves with corresponding FLAI, where each row corresponds to the ND, WA, and SA data anomaly class. Each light curve is standardised using its own mean and standard deviation.
}
\label{fig:kepler_sample}
\end{figure}

The second column of Table \ref{tab:proportions} summarises the proportion of Kepler data falling into the anomaly classes we defined above, based on the FLAI value. In particular, it shows that $36.2\%$ and $30.0\%$ of the data belong to the WA and SA classes, respectively. Three randomly selected examples of Kepler flare light curves corresponding to each data anomaly class are shown in Fig.~\ref{fig:kepler_sample}. In this example, the light curves from the ND class exhibit a clear \lq\lq standard\rq\rq\ flare shape contaminated by noise. The example flares from the WA class show visibly stronger fluctuations with higher modulation amplitudes compared to the ND case. The SA class light curves are found to contain several pronounced flare peaks, which makes them distinctly different from a \lq\lq standard\rq\rq\ flare.

The same dataset of 2,274 Kepler flare light curves was previously used to detect exponentially decaying QPP using an FCN approach \citep[see][Table 3]{Belov2024}. The FCN identified 159 events (7\% of the dataset) as exhibiting signatures of QPPs above a 95\% confidence threshold. We reconsidered those 159 Kepler flare events with our Deep SVDD model and obtained the FLAI values for them to fall into the ND class (35 events), WA class (55 events), and SA class (69 events). A corresponding table, containing the information about these events and their FLAI, can be found in Appendix \ref{sec:app}.  This illustrates the consistency between these two approaches: the majority of flare light curves, identified as QPP-containing by the FCN model, were identified as anomalous by the Deep SVDD model. The FCN model was trained to detect features specific to exponentially decaying QPPs, in contrast to the current model, which was trained only on noisy flare trends. This implies that the FCN can potentially identify QPPs even at higher $NSR$ values, whereas for the Deep SVDD model, the QPP and non-QPP data distributions increasingly overlap as $NSR$ increases (see Fig.~\ref{fig:train_dist}).

\begin{table}
\centering
\caption{Proportion of data in anomaly classes \label{tab:proportions}}
\begin{tabular}{ c c c c } 
 \hline
 Data type & Kepler  & STIX 4--10\,keV & STIX 15--25\,keV\\ 
 \hline
 Normal data (ND) & 769 (33.8\%) & 418 (70.0\%) & 287 (42.5\%)
\\
\hline
Weak anomalies (WA) & 822   (36.2\%) & 91 (15.2\%) & 172 (25.5\%)\\
\hline
Strong anomalies (SA)  & 683   (30.0\%) & 88 (14.8\%) & 216  (32.0\%)\\
\hline
\end{tabular}
\end{table}

\subsection{STIX flares}\label{sec:stix}

To proceed with solar flares, we used data from the Spectrometer/Telescope for Imaging X-rays \citep[STIX;][]{krucker2020} onboard Solar Orbiter \citep{mueller2020}. STIX measures solar hard X-ray emission in the energy range 4--150\,keV, providing diagnostics of both thermal plasma emission at lower energies and non-thermal bremsstrahlung emission from accelerated electrons at higher energies. STIX provides continuous solar monitoring, making it well suited for statistical studies of solar flare properties. We make use of the STIX quick-look light curves, which provide full-sun integrated count rates at a cadence of 4 seconds and are available for the full mission duration, making them ideal for large-scale statistical analysis of this kind. {While STIX is capable of providing science data at sub-second cadence, in this study we use the quick-look light curve data with a 4-second cadence, which are well suited to large-scale statistical analyses but may not capture the fastest temporal structures seen during the impulsive phase. The subsequent resampling of all flare profiles to 300 time bins introduces a further effective downsampling, particularly for the longest events, which can reduce sensitivity to short-timescale variability.} For this study, we used the STIX flare list\footnote{github.com/hayesla/stix\_flarelist\_science} containing flares occurred between 14 April 2021 and 28 February 2025 to extract STIX quick-look light curves. We limit our analysis to larger solar flare events, and filtered for flares that had an estimated GOES class of M- or X- (see \cite{xiao_2023} for how the GOES class estimates are achieved). In addition, we extract information only from the 4--10\,keV and 15--25\,keV energy bands, representing low- and high-energy bands, respectively. We filtered the data to exclude cases where the attenuator was used by limiting the peak flux in the low-energy band to $0.8\times10^5$ counts per second; we did not filter this way for the high-energy band, as it has a lower impact on a combined thermal and non-thermal emission seen in this channel.  We also select flares whose most prominent peaks deviate by more than five standard deviations from the mean calculated over the last 10\% of each light curve. Moreover, light curves without a pronounced rise phase are excluded. This selection procedure results in 597 and 675 flare events in the low- and high-energy bands, respectively. The dataset is available on Harvard Dataverse \citep{stix_dataset}.

The distributions of the data from both energy bands in terms of $d/R$, resulting from the application of our Deep SVDD model, are shown in the right panel of Fig.~\ref{fig:d_dist}. Both distributions are clearly visually separated, which indicates a statistical difference between the temporal morphology of flares seen in the low- and high-energy bands. This result can be associated with the different physical processes dominating emission in each channel. The 4--10\,keV band is dominated by thermal bremsstrahlung from hot flare plasma, and is therefore associated with the gradual phase of the flare. The 15--25\,keV band contains contributions from both thermal bremsstrahlung, which can dominate at these energies during the peak of large flares, and non-thermal bremsstrahlung from accelerated electrons, which tends to dominate during the impulsive phase. While the temporal profiles of thermal and non-thermal flare emission are related through the energy release and particle acceleration processes, they are known to differ morphologically, with non-thermal emission typically exhibiting a more impulsive and structured profile during the flare rise phase \citep{Fletcher2011, Benz2016}. The separation of the two distributions seen here is therefore consistent with these known physical differences, and suggests that analysing the full distribution rather than the average profile is a more sensitive way of revealing these morphological differences. In addition, the high-energy channel data contains a larger fraction of anomalous data according to Fig.~\ref{fig:d_dist}. For example, according to Table~\ref{tab:proportions}, 70.0\% of the low-energy data are classified as ND (with 15\% being WA and SA), whereas in high-energies only 42.5\% is classified as ND by the model. Moreover, the high-energy channel contains approximately twice as many flares classified as SA (32\% vs. 15\%). It should be mentioned that many light curves were downsampled to match the model input size of 300 data points, which could affect the model’s sensitivity to fast-evolving features. {It is also worth noting that the Kepler white-light distribution in Fig~\ref{fig:d_dist} lies closer to the STIX 15-25~keV distribution than to the 4-10~keV one. This is consistent with the long-established picture in which solar white-light continuum is produced, directly or indirectly, by the same non-thermal electron beams responsible for the impulsive hard X-ray emission \citep{Fletcher2007, watanabe_2010, kuhar_2016}. A more detailed physical discussion is given in Section~\ref{sec:concl}.}

\begin{figure}[h]
\centering
\includegraphics[width=0.8\textwidth]{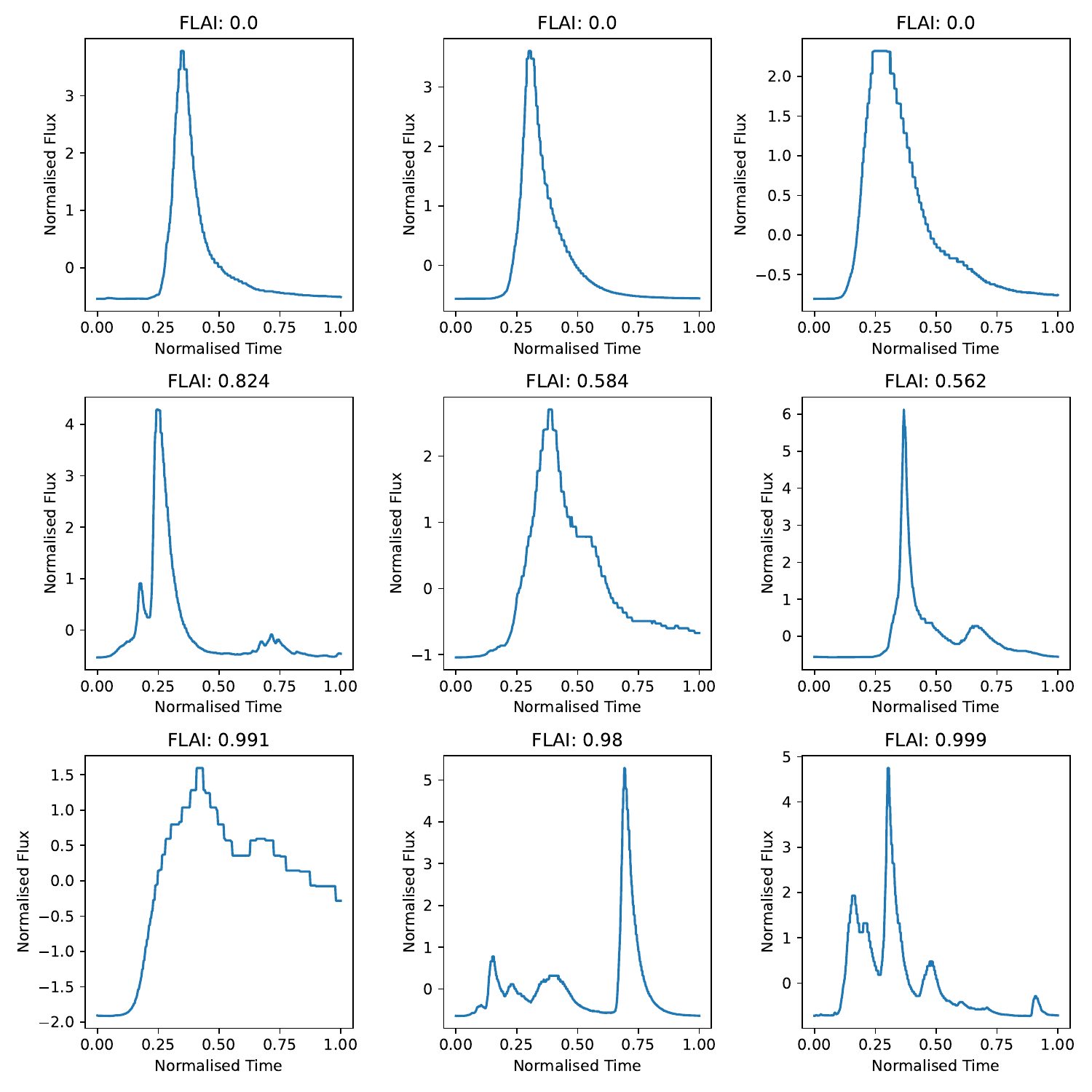}
\caption{Examples of STIX low-energy band (4-10 keV) light curves with corresponding FLAI, where each row corresponds to the ND, WA, and SA data anomaly class. Each light curve is standardised using its own mean and standard deviation.
}
\label{fig:stix_sample_low}
\end{figure}

\begin{figure}[h]
\centering
\includegraphics[width=0.8\textwidth]{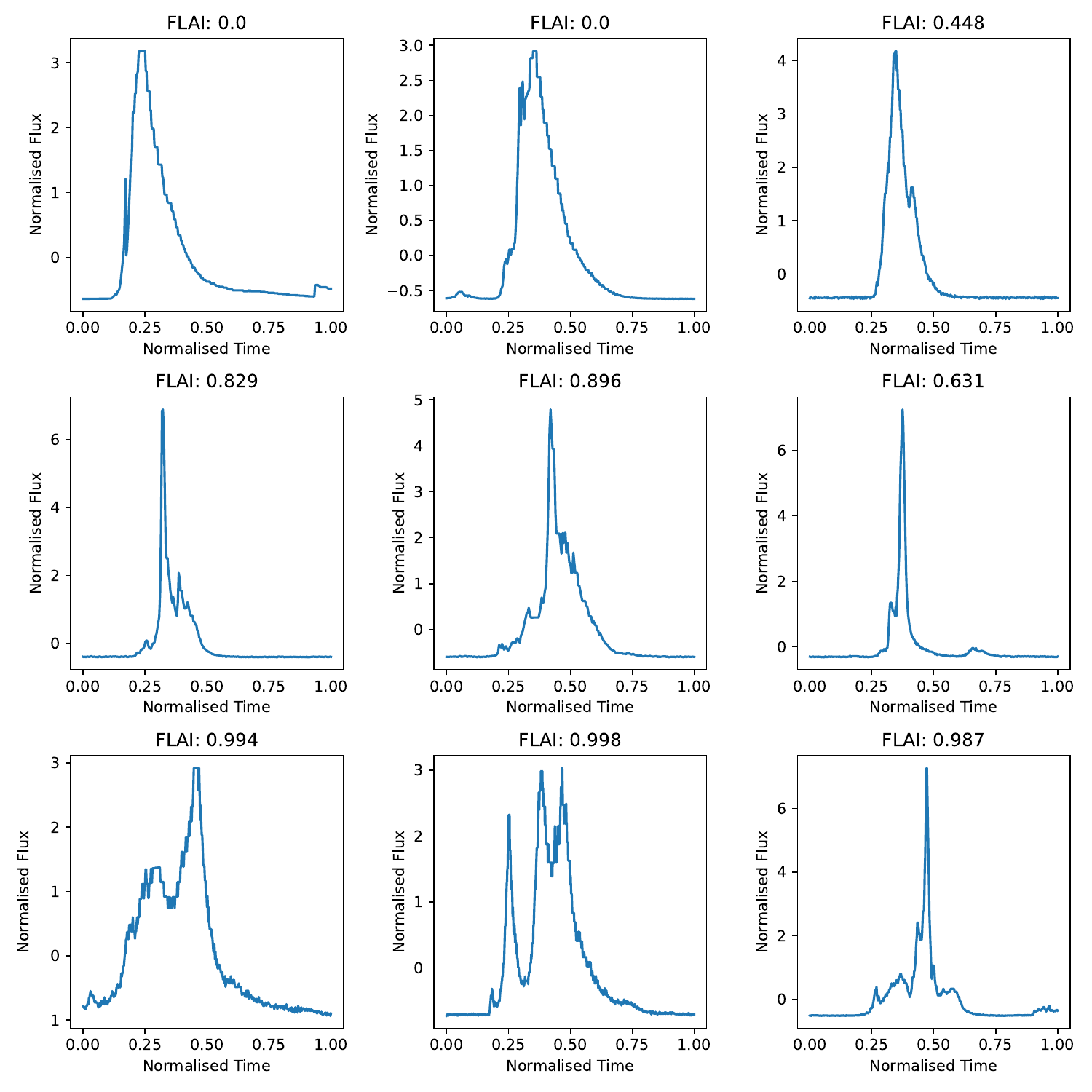}
\caption{Examples of STIX high-energy band (15-25 keV) light curves with corresponding FLAI, where each row corresponds to the ND, WA, and SA data anomaly class. Each light curve is standardised using its own mean and standard deviation.
}
\label{fig:stix_sample_high}
\end{figure}

Figure~\ref{fig:stix_sample_low} shows three randomly selected STIX light curves for every anomaly class in the low-energy band. The first row represents relatively smooth light curves with a regular flare shape. In the WA and SA classes, the example light curves show either a clear multi-peak structure or significantly distorted profiles. The bottom-left light curve from the SA class (with FLAI $=0.991$) is likely affected by an incorrectly estimated flare end time. Similar progressively increasing morphological complexity with multi-peak structure of SA samples is seen in the high-energy band, examples of which are shown in Fig.~\ref{fig:stix_sample_high}.

\subsection{Anomaly classification}
{The introduced anomaly-detection approach does not distinguish between different types of anomalies. In this sense, all anomalies identified by the model are statistical anomalies, as their statistical properties deviate from those of the training data. In real observations, these statistical anomalies correspond to both physically meaningful phenomena such as QPPs, multiple flares, and flares with unusual temporal profiles, and instrumental or data-processing artefacts (e.g. attenuator effects or imperfect trend removal). Consequently, both the WA and SA classes may contain a mixture of physical and instrumental/processing anomalies. Nevertheless, different anomaly types are expected to produce different FLAI distributions. For example, multiple flares are likely to yield systematically higher FLAI values than moderate-amplitude QPP signals. To investigate how FLAI is distributed among different anomaly classes, we used the Kepler and STIX low-energy-band datasets and randomly selected 50 flares from each of the WA and SA classes (100 events from each dataset) for further inspection. Then, we visually separated the data into 5 corresponding classes:
\begin{itemize}
    \item Multiple peaks/flares: events containing several overlapping or well-separated flares; 
    \item QPP: events exhibiting visually identifiable QPP behaviour; 
    \item Shape: flares whose temporal profiles differ significantly from the standard asymmetric flare shape and cannot be attributed to the other anomaly categories; 
    \item Instrumental/preprocessing artefacts: events affected by instrumental effects (e.g. attenuator-related distortions) or data-processing artefacts (e.g. poor trend removal); 
    \item Noise: events dominated by a visually high noise level. 
\end{itemize}
This classification scheme and the associated visual inspection are inherently subjective and may introduce biases, highlighting the need for additional analysis and more objective unsupervised classification techniques. For example, the distinction between QPP events and multi-peak flares is often ambiguous, and further analysis is required to demonstrate that the observed variability corresponds to genuine QPP behaviour rather than correlated noise. Nevertheless, this preliminary visual classification provides a useful first step for identifying potentially interesting subclasses of anomalies prior to the application of more rigorous and computationally intensive methods.}

\begin{figure}[h]
\centering
\includegraphics[width=\textwidth]{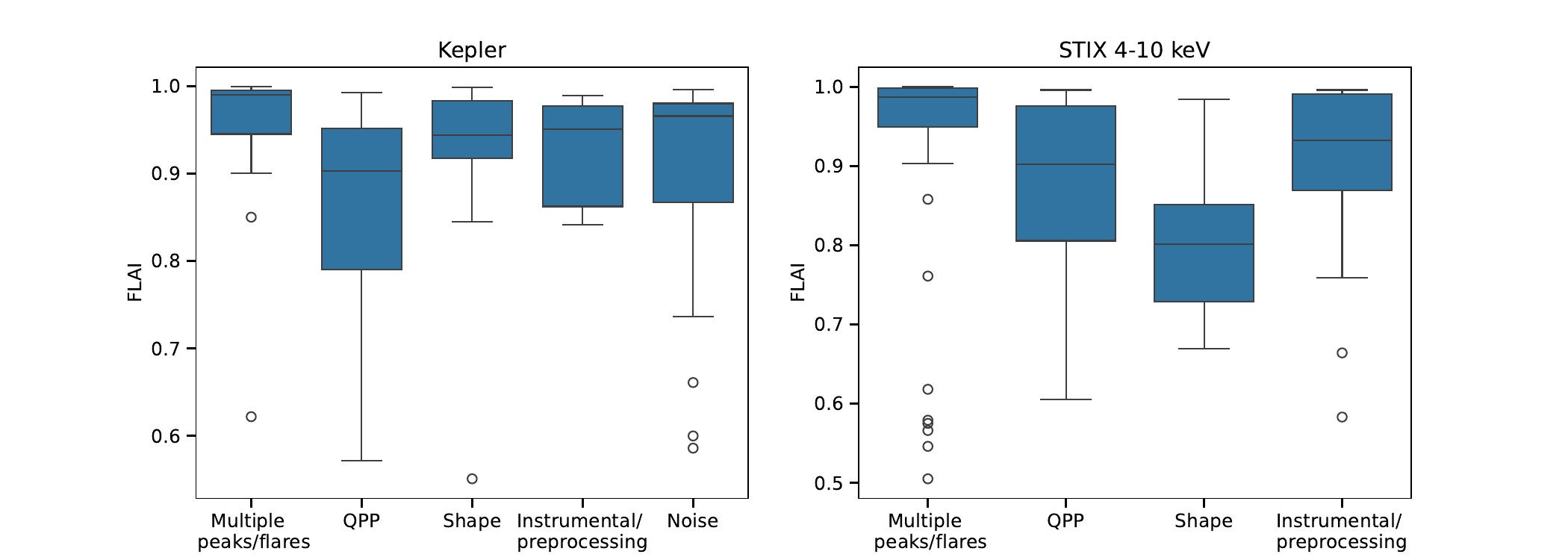}
\caption{{Box-and-whisker plots for anomaly classes from Kepler (left panel) and STIX low-energy band (right panel) data. The horizontal black line denotes a median value. Boundaries of the boxes are 25\% and 75\% levels, while whiskers are 5\% and 95\% levels. Black circles represent outliers for each class.}
}\label{fig:an_test}
\end{figure}

{Fig.~\ref{fig:an_test} shows the FLAI distributions for the proposed anomaly classes derived from the Kepler dataset (left panel) and the STIX low-energy-band dataset (right panel). Light curves containing multiple peaks or flares exhibit the highest FLAI values, with median values exceeding 0.95. In contrast, the QPP class displays the broadest FLAI distribution among all proposed categories. This behaviour is expected, as the class includes QPP signals spanning a wide range of amplitudes, which naturally produces a broad spread of FLAI values (see, for example, the right panel of Fig.~\ref{fig:f_test}). For the STIX data, the shape class exhibits the lowest median FLAI values. This is likely because the deviations associated with this class are relatively weak and affect the flare profile globally rather than producing strong localised anomalies. A similar trend is not observed in the Kepler data, which are characterised by a higher noise level. Instrumental and processing artefacts also produce systematically high FLAI values, with median values above 0.9. Finally, the noise class in the Kepler dataset shows the second broadest FLAI distribution after the QPP class, whereas the STIX data are not sufficiently noisy for such a category to emerge clearly. Example light curves with normalised integrated-gradient attributions \citep{pmlr-v70-sundararajan17a} for each anomaly class are presented in Appendix~\ref{sec:app2}.}

\section{Do anomalous flares have a template?} \label{sec:meanprofile}

In previous studies of flare templates \citep{Davenport2014, Kashapova2021, Motyk2025}, the authors deliberately focused on subsets of flares resembling a standard profile, i.e. characterised by a rapid rise phase and a slow decay phase without secondary peaks or bursts. Such selection was generally performed through visual inspection of flare light curves and may therefore have implicitly included flares with more complex temporal morphologies (see, e.g., Figs.~1--3 in \cite{Kashapova2021} and Fig.~3 in \cite{Motyk2025}, which illustrate significant variability of the data around the average trend). In the context of the present work, this raises the question of whether a single representative flare template remains a meaningful descriptor for increasingly anomalous flares.

\begin{figure}[h]
\centering
\includegraphics[width=0.32\textwidth]{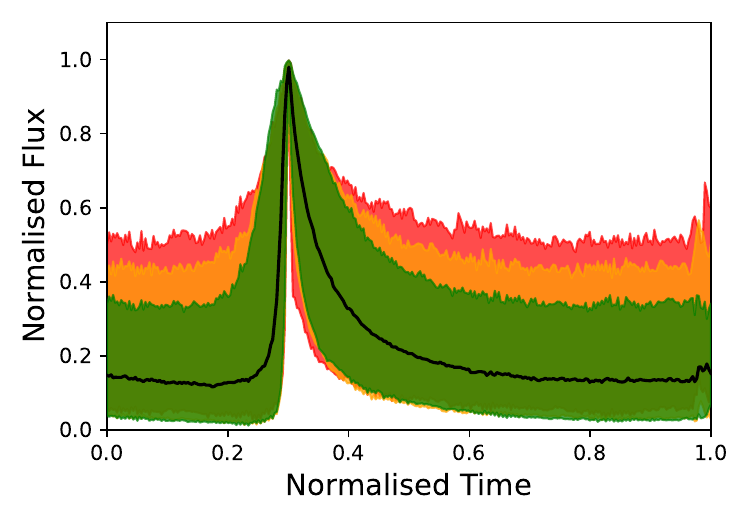}
\includegraphics[width=0.32\textwidth]{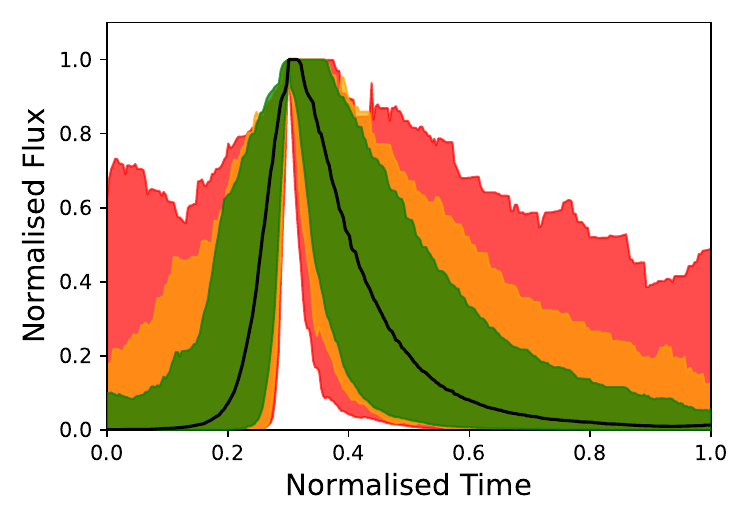}
\includegraphics[width=0.32\textwidth]{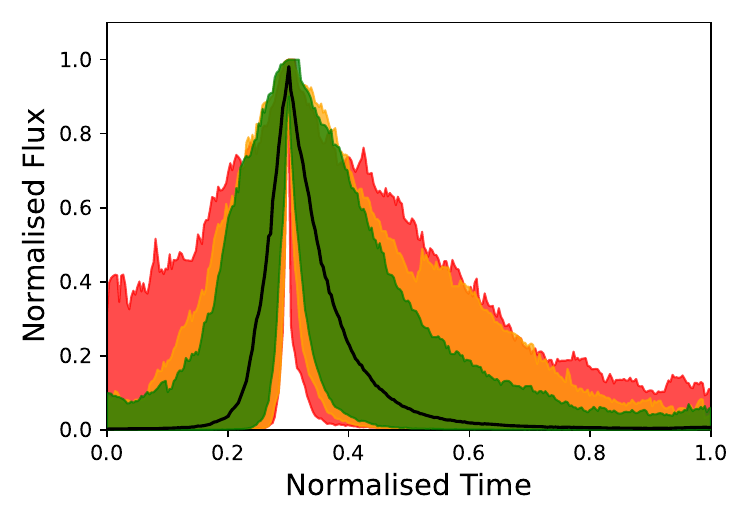}
\caption{Median flare profile in ND (black curve) and the data regions between 5- and 95- percentiles for data from Kepler (left column), STIX low-energy (middle column) high-energy (right column) bands seperated into ND (green region), WA (orange region), and SA (red region) classes.
}
\label{fig:med_profiles}
\end{figure}

To examine flare templates and variability within each anomaly class, we normalised each flare duration to unity, applied min–max scaling to the flare profiles, and aligned the profiles by their peak times for both the Kepler and STIX datasets. Using these processed light curves, we calculated the median, as well as the 5- and 95- percentiles, in each temporal bin. These results are shown in Fig.~\ref{fig:med_profiles}. We found that the median profiles for the ND, WA, and SA classes are very similar; therefore, only the median profile for the ND class is shown. However, the spread of the flare profiles around this median differs substantially between the three classes and increases systematically with anomaly level. The green, orange, and red shaded regions enclose 90\% of the flare data (between the 5- and 95- percentiles) for the ND, WA, and SA classes, respectively. It is evident that flare variability increases with anomaly class. For example, for the STIX low-energy band at $t = 0.5$, the coefficient of variation (the difference between the 95th percentile and the flare median, normalised to the latter) increases from 1.8 in the ND class to 2.3 and 3.1 in the WA and SA classes, respectively. This indicates that increasingly anomalous flares deviate more strongly from the median flare template, even if their median profile remains broadly unchanged. In other words, anomaly level is not primarily associated with a change in the average flare shape, but rather with a reduction in how well a single common template represents the population. Consequently, defining a single representative flare template for the more anomalous classes may be of limited value, as these classes likely comprise a diverse set of distinct flare morphologies that may require an individual treatment.

\section{Conclusions} 
\label{sec:concl}
In this study, we used a Deep SVDD approach \citep{pmlr-v80-ruff18a} and the FCN architecture \citep{Wang2017} to identify solar and stellar flares deviating from a \lq\lq typical\rq\rq\ flare profile. As a model of a \lq\lq typical\rq\rq\ flare profile, we used 45,000 flare trends flare trends derived from stellar \citep{Davenport2014} and solar flare trend models \citep{Gryciuk2017, Broomhall2019} taken from the dataset \citep{qpp_dataset} previously used to train a neural network for distinguishing flares with and without QPPs \citep{Belov2024}. Then, the model was trained on this dataset containing only \lq\lq typical\rq\rq\ flare profiles. To prevent overfitting, we introduced noise dynamically during training, with a noise-to-signal ratio between 0.01 and 0.1, ensuring that the model did not encounter the same light curve twice. Additionally, to regularise the model, we used \lq\lq soft\rq\rq\ boundaries allowing 10\% of the normal data to lie outside the hypersphere, while encouraging the model to learn a compact representation of the core data distribution. This corresponds to an unsupervised learning approach, meaning that the model is trained without being provided with explicit \lq\lq correct answers\rq\rq\ and instead learns patterns and relationships directly from the data. This distinguishes our method from previous studies \citep{Vida2021, Jia2024, Belov2024, Tan2025}, where models were trained in a supervised manner using labeled datasets.

Before applying the developed model to real light curves, we introduced Flare Anomaly Index (FLAI) as a measure of how much a particular light curve deviates from the normal data. This index is determined by Equation (\ref{eq:flai}) and quantifies the probability of a light curve being anomalous. To calculate FLAI, we estimated the probability of a given event belonging to the tail of the normal data distribution by fitting the tail with a generalised Pareto distribution (GPD). The  hypersphere radius $R$ was used as the threshold value to define the start of the distribution tail. We then divided the data into three classes based on the FLAI value: normal data (ND) for $0.0\le FLAI<0.5$, weak anomalies (WA) for $0.5\le FLAI<0.95$, and strong anomalies (SA) for $FLAI\ge0.95$.

We then applied the method to light curves from the Kepler stellar flare catalogue \citep{Balona2015} to calculate the corresponding FLAI values. The distribution of the data across anomaly classes and representative examples is presented in Table \ref{tab:proportions} and Fig. \ref{fig:kepler_sample}, respectively. For this dataset, the data are approximately evenly distributed among the ND, WA, and SA classes. We repeated the same procedure for STIX quick-look light curves selected from flares occurring between 14 April 2021 and 28 February 2025, with an estimated GOES class of M or X. We used the data from 4-10 keV and 15-25 keV bands representing low- and high- energy channels, respectively. The results are summarised in Table~\ref{tab:proportions}. In particular, $70\%$ of the low-energy band data are classified as normal, whereas only 42.5\% fall into the normal class in the high-energy band. The proportion of data in the WA and SA classes for the high-energy band is approximately twice as large as that for the low-energy band. Representative light curves from each class are shown in Figs.~\ref{fig:stix_sample_low} and \ref{fig:stix_sample_high} for the 4-10 keV and 15-25 keV bands, respectively. However, these results may be affected by instrumental artefacts, including attenuator effects, as well as by the presence of multiple flares within a single time series due to limitations of the STIX flare list, and data downsampling. Future studies focusing on flare anomalies should therefore incorporate improved data cleaning to minimise these effects. In addition, we calculated the median, as well as the 5- and 95- percentiles, for each anomaly class in both the Kepler and STIX datasets. We found that the median profiles for the ND, WA, and SA classes are very similar, while flare variability increases systematically with anomaly class, which makes the concept of a typical flare profile less meaningful for anomalous flares.

We made the model publicly available by publishing it together with the training notebook and a web application on GitHub\footnote{https://github.com/Warwick-Solar/FLAI} and Zenodo \citep{FLAI}. Following the practice we employ for the ML-tools \citep{Belov2024, Belov2025}, we created an interactive tool to use the developed model. The source code and the instructions on how to install it can be found in the above repositories. 

The introduced method has several important implications. First, it allows the analysis of data distributions for large ensembles and enables comparisons between datasets from different sources. In particular, Fig.~\ref{fig:d_dist} demonstrates that the Kepler white-light data distribution is closer to that of the non-thermal STIX channel than to the thermal one. {This is consistent with the standard picture in which solar white-light continuum is powered by the same non-thermal electron beams responsible for impulsive hard X-ray emission, either through direct collisional heating of the chromosphere \citep{Fletcher2007} or through indirect radiative back-warming of the lower atmosphere \citep{Machado1989, allred_2005}. Previous observational studies also support such a connection showing a close spatial and temporal correspondence between white-light and hard X-ray emission, which has been confirmed both by individual simultaneous observations \citep{watanabe_2010, MartnezOliveros2012} and by statistical studies of M- and X--class flares observed jointly by SDO/HMI and RHESSI \citep{kuhar_2016, Huang2016}.
For example, \cite{kuhar_2016} performed a statistical analysis of 43 M- and X-class flares observed simultaneously by SDO/HMI (white light) and RHESSI (hard X-ray) and found a clear positive correlation between the hard X-ray flux at 30~keV and the white-light flux at 6173~\AA, integrated around the hard X-ray peak. In particular, white-light emission was detected in essentially all flares with sufficiently high hard X-ray fluxes, and the deposited non-thermal electron energy was found to be sufficient to power the observed white-light losses. These results support the picture in which white-light and hard X-ray emission share a common origin in the impulsive energy release, while still depending on different emission and atmospheric response mechanisms, such as direct electron-beam heating, chromospheric radiative back-warming, and continuum formation at different atmospheric heights \citep{Machado1989, Fletcher2007, Battaglia2011}.  We note, however, that the solar white-light/hard X-ray analogy does not transfer perfectly to the stellar regime: M-dwarf flares in particular show evidence for an additional hot ($T \sim 10^4$~K) blackbody-like continuum component whose origin and temporal evolution may differ from the prompt hard X-ray response \citep{Kowalski2013, kowalski_2016}, and we refer the reader to \cite{Kowalski2024} for a comprehensive recent review of solar and stellar flare continuum emission.}
Moreover, the anomalous flares identified in this study show strong deviations from the median flare template, indicating that a single common template provides a poor representation of the population. This further suggests that highly anomalous flares may be less amenable to simple parametric characterisation in terms of a single rise time, decay time, or flare duration, and may instead reflect compound or temporally structured energy-release processes. In addition, the introduced FLAI demonstrates that QPPs form a subset of a broader class of flare anomalies. Moreover, it is a promising avenue to investigate how flare energy or class relates to anomalous behaviour. In this context, it is particularly interesting to explore whether rare and powerful events such as super flares \citep{Maehara2012, Shibayama2013, Namekata2017, Vasilyev2024} exhibit anomalous temporal morphology.

It should be mentioned that the results presented are subject to a specific choice of the normal-data model and how well it represents typical features of the events. It implies that the data model can be adjusted for the task. First, the physical model of a typical event can be modified to account for the physical mechanisms of interest. {In particular, the current data model assumes single flare profiles prescribed by analytical and phenomenological models \citep{Davenport2014, Gryciuk2017, Broomhall2019} with parameters chosen randomly from a broad distribution. In contrast, in real flares, these parameters can be correlated and follow a different distribution. For example, temperature- and density-dependent conduction and radiative losses can modify the decay-phase slope and duration for more energetic flares \citep{Cargill1995, 2025ApJ...987L...9B}. Thus, including physically motivated scalings/laws in the data model can help identify events that do not obey them. In particular, the data model can include two flare channels following the Neupert effect \citep{Neupert1968} to find flares significantly deviating from it \citep{Dennis1993, Tristan2023}.} On the other hand, instrument-specific adjustments can be implemented. {The instrument cadence and the typical durations of the events seen by a specific instrument can be accounted for during the training data generation to improve the data quality. Moreover, the account for the spectral properties of noise observed by a specific instrument can potentially reduce the domain gap between synthetic and real datasets. Also, previously identified instrumental artefacts can be included in the training data to filter them from real physical anomalies.} Thus, it allows one to fine-tune the presented technique for a given problem and instrument {and limit the anomaly type under consideration}.

Thus, the developed model can enhance flare analysis pipelines. This can be achieved either through its standalone application or by combining it with other techniques, including deep-learning models for flare detection \citep{Jia2024}, QPP-detection tasks \citep{Belov2024}, and transform-based approaches \citep{Broomhall2019}.

\begin{acknowledgments}
The work is supported by the STFC Grant ST/X000915/1 and the Latvian Science Council Grant lzp-2024/1-0023. DYK also thanks the UKRI Stephen
Hawking Fellowship EP/Z535473/1. L.A.H is supported by a Royal Society-Research Ireland University Research Fellowship (URF$\backslash$R1$\backslash$241775).
\end{acknowledgments}

\software{NumPy, a Python package for fundamental scientific computing \citep{harris2020array};
SciPy, a Python package for fundamental algorithms for scientific computing \citep{2020SciPy-NMeth}; 
PyTorch Lightning, the deep learning framework; Streamlit, a Python library for deploying ML projects; STIXpy, a Python package for analysing data from the Spectrometer/Telescope for Imaging X-rays (STIX) instrument onboard the Solar Orbiter spacecraft \citep{stixpy}.}

%






\appendix

\section{FLAI for Kepler flares with identified QPPs}
\label{sec:app}

\begin{longtable}{ccc|ccc|ccc}
\caption{Kepler flare events with identified presence of QPPs \citep{Belov2024} and their FLAI. The corresponding light curves can be found in \citet{kepler_dataset}.} \\

\endfirsthead
\endhead
\endfoot

\multicolumn{9}{l}{\textbf{Normal Data}} \\
\addlinespace[0.3em]
\toprule
KIC & BJD & FLAI & KIC & BJD & FLAI & KIC & BJD & FLAI \\
\midrule

11560431 & 2456057.61 & 0.0 & 12156549 & 2455086.645 & 0.0 & 10528093 & 2456262.771 & 0.0 \\
11560431 & 2455013.574 & 0.0 & 8429280 & 2455006.133 & 0.0 & 11560431 & 2455010.158 & 0.0 \\
11551430 & 2456170.457 & 0.0 & 11551430 & 2456235.537 & 0.0 & 11560431 & 2456142.847 & 0.0 \\
7339343 & 2455005.364 & 0.0 & 7940546 & 2455048.776 & 0.0 & 4831454 & 2455187.025 & 0.0 \\
11560431 & 2456117.885 & 0.0 & 10063343 & 2455151.39 & 0.0 & 11560431 & 2456172.309 & 0.0 \\
11231334 & 2455491.728 & 0.0 & 4543412 & 2455163.976 & 0.0 & 2300039 & 2455773.901 & 0.0 \\
6548447 & 2455323.093 & 0.0 & 4671547 & 2455087.968 & 0.0 & 11560431 & 2456165.778 & 0.0 \\
11551430 & 2456278.089 & 0.0 & 11610797 & 2454981.631 & 0.0 & 9821078 & 2455705.652 & 0.0 \\
11560431 & 2456056.445 & 0.0 & 10355856 & 2454982.87 & 0.0 & 11560431 & 2456114.382 & 0.0 \\
5733906 & 2455099.532 & 0.0 & 11560431 & 2456070.227 & 0.0 & 11551430 & 2456296.8 & 0.0 \\
11551430 & 2455029.862 & 0.0 & 11551430 & 2456217.422 & 0.0 & 11560431 & 2455023.444 & 0.144 \\
11665620 & 2455791.997 & 0.294 & 7940546 & 2456259.122 & 0.37 &  &  & \\

\addlinespace

\multicolumn{9}{l}{\textbf{Weakly Anomalous}} \\
\addlinespace[0.3em]
\toprule
KIC & BJD & FLAI & KIC & BJD & FLAI & KIC & BJD & FLAI \\
\midrule

12102573 & 2455072.339 & 0.506 & 7206837 & 2456084.741 & 0.611 & 10160534 & 2455064.852 & 0.617 \\
11709006 & 2454966.828 & 0.637 & 12156549 & 2455342.383 & 0.648 & 4758595 & 2456226.538 & 0.691 \\
11560431 & 2456173.104 & 0.702 & 4758595 & 2456233.845 & 0.71 & 5522786 & 2455871.3 & 0.713 \\
11560431 & 2456197.291 & 0.733 & 12102573 & 2455079.331 & 0.735 & 5609753 & 2456399.539 & 0.736 \\
7206837 & 2455806.894 & 0.737 & 5557932 & 2455038.978 & 0.746 & 3239945 & 2456170.505 & 0.75 \\
4671547 & 2455090.038 & 0.766 & 5357275 & 2455106.976 & 0.78 & 5557932 & 2455039.036 & 0.785 \\
3441906 & 2455952.2 & 0.788 & 10355856 & 2455387.912 & 0.79 & 11560431 & 2456156.548 & 0.793 \\
5108214 & 2455837.489 & 0.796 & 7940546 & 2455871.277 & 0.804 & 6106152 & 2455013.524 & 0.808 \\
11551692 & 2456289.977 & 0.824 & 12156549 & 2455347.197 & 0.837 & 9641031 & 2455022.582 & 0.841 \\
8429280 & 2455032.246 & 0.85 & 4671547 & 2455073.743 & 0.855 & 11560431 & 2456132.526 & 0.858 \\
11560431 & 2456170.503 & 0.86 & 11548140 & 2455486.206 & 0.864 & 11560447 & 2455055.708 & 0.89 \\
4273689 & 2455249.467 & 0.895 & 11560431 & 2456064.381 & 0.899 & 7940546 & 2455768.88 & 0.901 \\
10459987 & 2456179.644 & 0.905 & 9349698 & 2456354.714 & 0.907 & 11709006 & 2455408.873 & 0.908 \\
11231334 & 2456178.25 & 0.913 & 4939265 & 2456281.794 & 0.913 & 2300039 & 2455790.849 & 0.914 \\
11551430 & 2456194.753 & 0.915 & 11560431 & 2456150.678 & 0.915 & 7841024 & 2454968.701 & 0.918 \\
4671547 & 2455074.027 & 0.93 & 3430868 & 2455039.036 & 0.938 & 7940546 & 2456094.427 & 0.939 \\
9705459 & 2456371.072 & 0.942 & 12156549 & 2455376.074 & 0.943 & 9641031 & 2456100.371 & 0.945 \\
7765135 & 2455072.194 & 0.945 & 11551430 & 2456245.555 & 0.947 & 3239945 & 2456156.116 & 0.947 \\
7940546 & 2455779.026 & 0.948 &  &  & &  & \\

\addlinespace

\multicolumn{9}{l}{\textbf{Strongly Anomalous}} \\
\addlinespace[0.3em]
\toprule
KIC & BJD & FLAI & KIC & BJD & FLAI & KIC & BJD & FLAI \\
\midrule

5475645 & 2456330.43 & 0.951 & 11551430 & 2455013.017 & 0.951 & 1025986 & 2455133.355 & 0.954 \\
3430868 & 2455038.978 & 0.955 & 9895004 & 2455437.151 & 0.959 & 11560431 & 2456112.473 & 0.959 \\
11560431 & 2455020.996 & 0.962 & 3128488 & 2454964.786 & 0.962 & 2162635 & 2455806.895 & 0.965 \\
12156549 & 2455067.352 & 0.967 & 11709006 & 2454971.707 & 0.967 & 12156549 & 2455374.593 & 0.967 \\
4568729 & 2455056.284 & 0.968 & 12418816 & 2455257.868 & 0.971 & 11560431 & 2455003.17 & 0.972 \\
11709006 & 2454969.451 & 0.974 & 4758595 & 2456210.581 & 0.974 & 9821078 & 2455623.573 & 0.975 \\
5357275 & 2455119.591 & 0.975 & 9655129 & 2456112.973 & 0.976 & 11551692 & 2456042.642 & 0.976 \\
11231334 & 2455435.206 & 0.977 & 9833666 & 2456423.294 & 0.979 & 6286925 & 2455038.42 & 0.979 \\
11548140 & 2455488.308 & 0.98 & 11560431 & 2456117.449 & 0.981 & 9833666 & 2456394.052 & 0.982 \\
11560431 & 2456115.361 & 0.982 & 2302548 & 2456170.505 & 0.982 & 11551430 & 2455024.522 & 0.982 \\
7692454 & 2456266.216 & 0.983 & 4758595 & 2456219.141 & 0.983 & 11548140 & 2455453.673 & 0.984 \\
6205460 & 2455334.886 & 0.984 & 1871056 & 2456156.115 & 0.985 & 9641031 & 2456219.273 & 0.986 \\
1161345 & 2456170.506 & 0.987 & 6106152 & 2455018.608 & 0.988 & 8429280 & 2455032.148 & 0.988 \\
11560431 & 2456110.758 & 0.989 & 6106152 & 2455024.179 & 0.99 & 9655129 & 2456148.63 & 0.99 \\
1161345 & 2455806.895 & 0.99 & 7940546 & 2455503.213 & 0.991 & 8651471 & 2455845.632 & 0.991 \\
9761199 & 2455799.706 & 0.991 & 12156549 & 2455346.612 & 0.992 & 11560431 & 2456184.967 & 0.992 \\
8429280 & 2455004.753 & 0.992 & 11560431 & 2456194.578 & 0.993 & 11189959 & 2455114.891 & 0.993 \\
11548140 & 2455386.295 & 0.993 & 11560431 & 2456186.81 & 0.994 & 11560431 & 2456072.751 & 0.994 \\
9641031 & 2456157.95 & 0.994 & 12156549 & 2455318.629 & 0.995 & 11709006 & 2454971.877 & 0.995 \\
11665620 & 2455806.894 & 0.995 & 9641031 & 2456109.318 & 0.995 & 11551692 & 2456170.504 & 0.995 \\
11709006 & 2455430.144 & 0.996 & 11551430 & 2456279.936 & 0.996 & 4273689 & 2455263.655 & 0.996 \\
5522786 & 2456011.854 & 0.997 & 9652680 & 2455090.13 & 0.997 & 6442183 & 2455860.26 & 0.997 \\
12644769 & 2456170.501 & 0.997 & 6106152 & 2455033.037 & 0.997 & 11551430 & 2456131.799 & 0.998 \\

\bottomrule
\end{longtable}

\section{Examples of anomaly classes}\label{sec:app2}
\begin{figure}[h]
\centering
\includegraphics[width=0.8\textwidth]{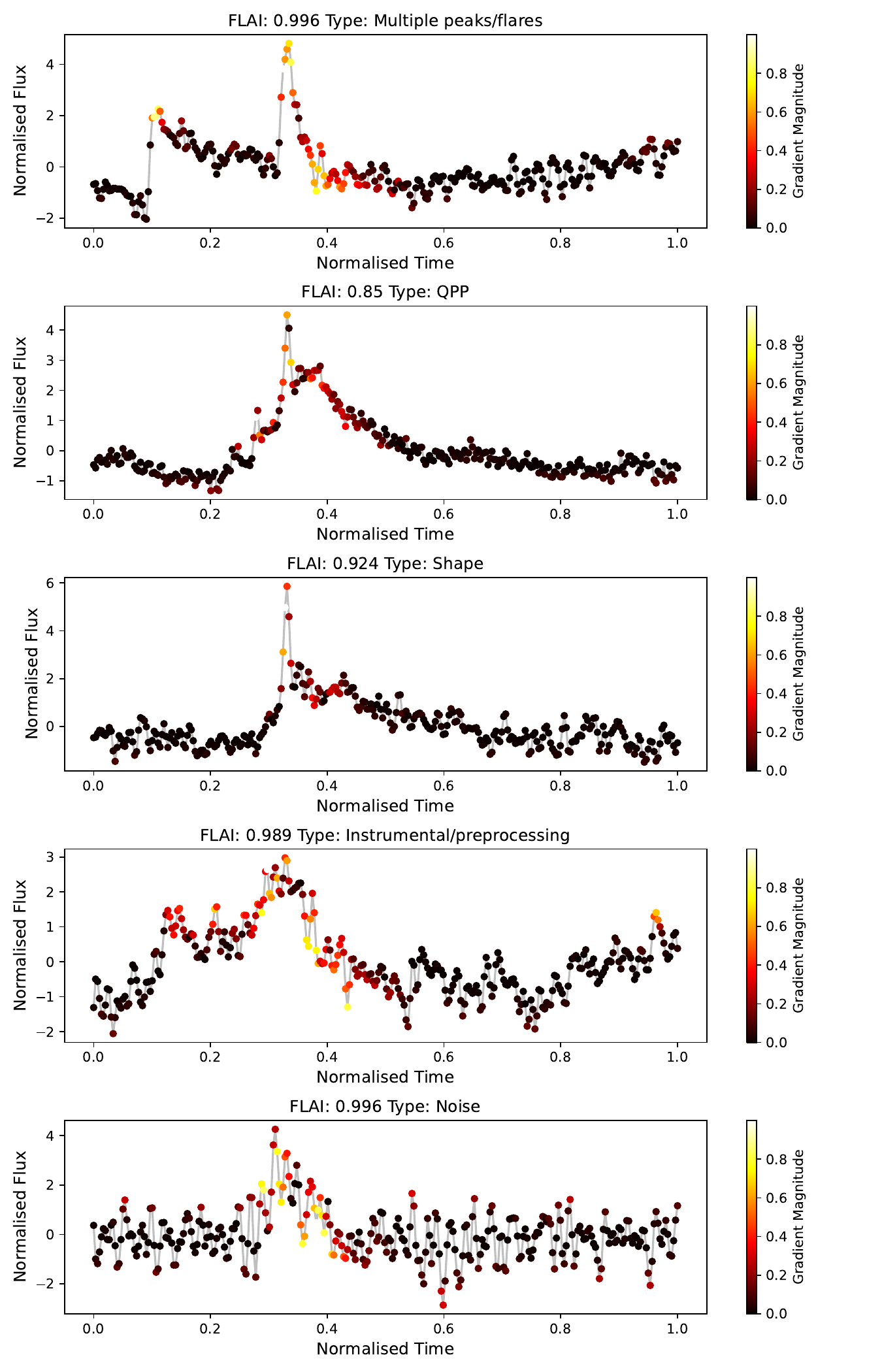}
\caption{{Examples of Kepler light curves with corresponding FLAI values and anomaly types. The colour scale shows the normalised integrated-gradient attribution, indicating the relative contribution of individual time points to the anomaly score. Each light curve is standardised by its own mean and standard deviation.}
}
\label{fig:app_examples0}
\end{figure}

\begin{figure}[h]
\centering
\includegraphics[width=0.8\textwidth, trim=0cm 5cm 0cm 0cm]{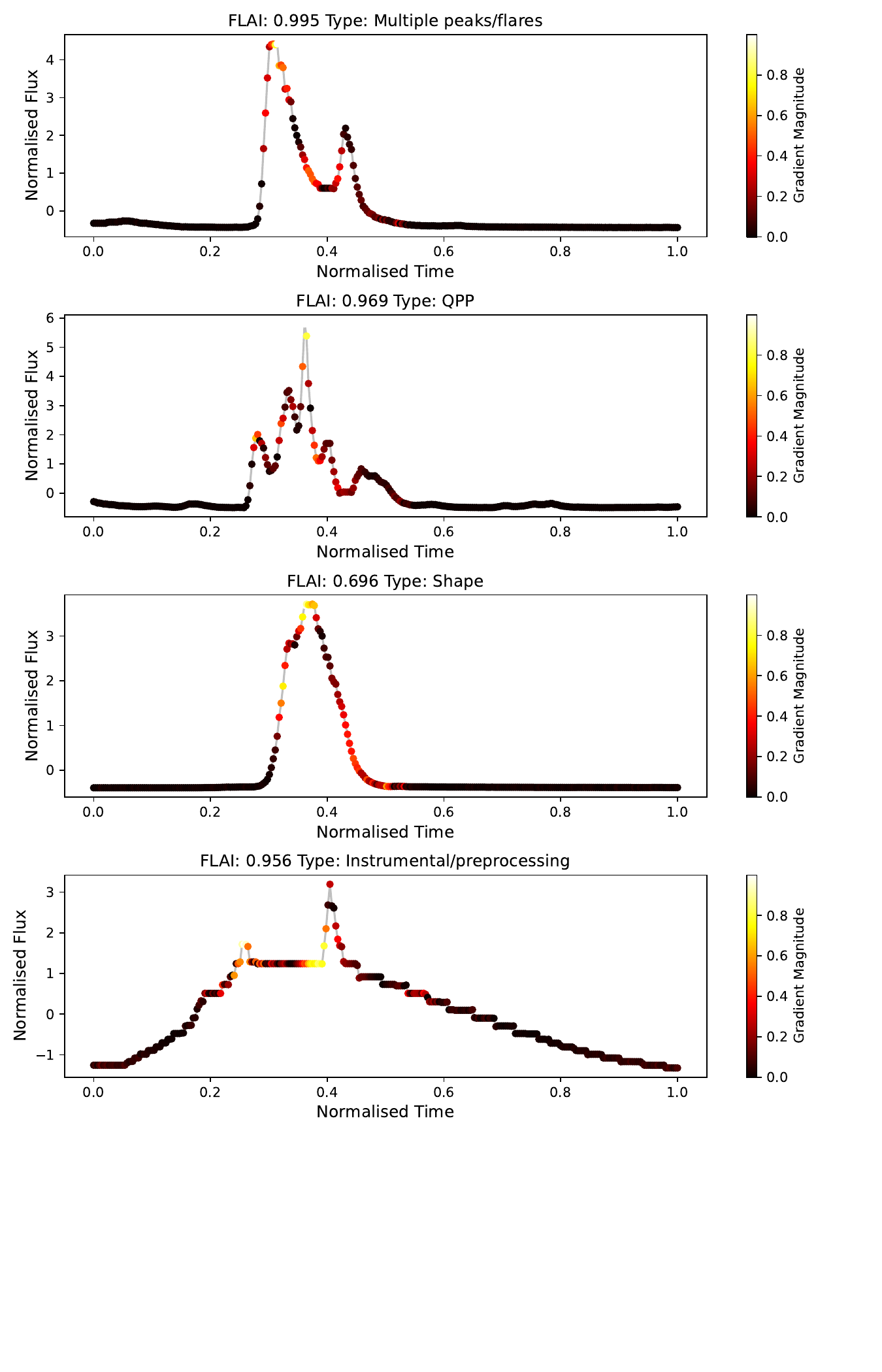}
\caption{{Examples of STIX low-energy band (4-10 keV) light curves with corresponding FLAI values and anomaly types. The colour scale shows the normalised integrated-gradient attribution, indicating the relative contribution of individual time points to the anomaly score. Each light curve is standardised by its own mean and standard deviation.}
}
\label{fig:app_examples1}
\end{figure}


\bibliographystyle{aasjournal}



\end{document}